\documentclass{aastex62}

\newcommand\rah{\mbox{$^{\mathrm h}$}}%
\newcommand\ram{\mbox{$^{\mathrm m}$}}%

\received{}
\revised{}
\accepted{}
\submitjournal{ApJL}

\shorttitle{A turnover in the radio light curve of GW170817}
\shortauthors{Dobie et al.}

\begin{document}

\title{A turnover in the radio light curve of GW170817}

\correspondingauthor{Tara Murphy}
\email{tara.murphy@sydney.edu.au}

\author[0000-0003-0699-7019]{Dougal Dobie}
\affiliation{Sydney Institute for Astronomy, School of Physics, University of Sydney, Sydney, New South Wales 2006, Australia.}
\affiliation{ARC Centre of Excellence for All-sky Astrophysics (CAASTRO), Australia.}
\affiliation{ATNF, CSIRO Astronomy and Space Science, PO Box 76, Epping, New South Wales 1710, Australia}

\author[0000-0001-6295-2881]{David L. Kaplan}
\affiliation{Department of Physics, University of Wisconsin - Milwaukee, Milwaukee, Wisconsin 53201, USA.}

\author[0000-0002-2686-438X]{Tara Murphy}
\affiliation{Sydney Institute for Astronomy, School of Physics, University of Sydney, Sydney, New South Wales 2006, Australia.}
\affiliation{ARC Centre of Excellence for All-sky Astrophysics (CAASTRO), Australia.}

\author[0000-0002-9994-1593]{Emil Lenc}
\affiliation{Sydney Institute for Astronomy, School of Physics, University of Sydney, Sydney, New South Wales 2006, Australia.}
\affiliation{ARC Centre of Excellence for All-sky Astrophysics (CAASTRO), Australia.}

\author[0000-0002-2557-5180]{Kunal P. Mooley}
\affil{National Radio Astronomy Observatory, Socorro, New Mexico 87801, USA}
\affil{Caltech, 1200 E. California Blvd. MC 249-17, Pasadena, CA 91125, USA}
\altaffiliation{Jansky Fellow (NRAO/Caltech).}

\author[0000-0002-0494-192X]{Christene Lynch}
\affiliation{Sydney Institute for Astronomy, School of Physics, University of Sydney, Sydney, New South Wales 2006, Australia.}
\affiliation{ARC Centre of Excellence for All-sky Astrophysics (CAASTRO), Australia.}

\author[0000-0001-8104-3536]{Alessandra Corsi}
\affiliation{Department of Physics and Astronomy, Texas Tech University, Box 41051, Lubbock, TX 79409-1051, USA}

\author{Dale Frail}
\affiliation{National Radio Astronomy Observatory, Socorro, New Mexico 87801, USA}

\author[0000-0002-5619-4938]{Mansi Kasliwal}
\affiliation{Caltech, 1200 E. California Blvd. MC 249-17, Pasadena, CA 91125, USA}

\author[0000-0002-7083-4049]{Gregg Hallinan}
\affiliation{Caltech, 1200 E. California Blvd. MC 249-17, Pasadena, CA 91125, USA}

\begin{abstract}
We present 2--9~GHz radio observations of \object{GW170817} covering the period 125--200 days post-merger, taken with the Australia Telescope Compact Array and the Karl G. Jansky Very Large Array. Our observations demonstrate that the radio afterglow peaked at $149\pm2$ days post-merger and is now declining in flux density.  We see no evidence for evolution in the radio-only spectral index, which remains consistent with optically-thin synchrotron emission connecting the radio, optical, and X-ray regimes.   The peak  implies a total energy in the synchrotron-emitting component of a ${\rm few}\times 10^{50}$\,erg. The temporal decay rate is most consistent with mildly- or non-relativistic material and we do not see evidence for a very energetic off-axis jet, but we cannot distinguish between a lower-energy jet and more isotropic emission. 
\end{abstract}


\section{Introduction} \label{sec:intro}
The neutron star merger \object{GW170817} was detected via the concurrent observation of gravitational waves \citep{2017PhRvL.119p1101A} and a $\gamma$-ray burst \citep[GRB;][]{2017ApJ...848L..13A,2017ApJ...848L..14G}. The merger was localized to its host galaxy, \object{NGC 4993}, by the detection of an optical transient \citep{2017Natur.551...64A,2017ApJ...848L..12A,2017Sci...358.1556C,2017ApJ...850L...1L,2017ApJ...848L..16S,2017ApJ...848L..27T,2017ApJ...848L..24V} and subsequent ultraviolet, optical and infrared observations found evidence of kilonova emission from the source \citep{2017Natur.551...64A,2017ApJ...848L..17C,2017Sci...358.1570D,Evans1565,2017Sci...358.1559K}. X-ray observations found no evidence of emission until nine days post-merger \citep{2017ApJ...848L..25H,Evans1565,2017ApJ...848L..20M,2017Natur.551...71T}, suggesting that this event differs significantly from previously-observed GRBs.

Radio emission from \object{GW170817} was first detected 16 days post-merger \citep{2017Sci...358.1579H}. Follow-up observations over the next 100\,days \citep{2017ApJ...848L..21A,2018arXiv180103531M,2018Natur.554..207M,2018arXiv180106516T} revealed a gradually rising light curve. The observed radio emission follows a power-law with temporal index $\delta=0.78\pm 0.05$ and spectral index $\alpha=-0.61\pm 0.05$, where $S_\nu(t,\nu) \propto t^\delta \nu^\alpha$ \citep{2018Natur.554..207M}. The observed radio spectral energy distribution agrees with the spectral index connecting contemporaneous radio, optical, and X-ray measurements, implying a common source for the observed synchrotron emission \citep{GCN22207,GCN22211,2018arXiv180103531M,2018Natur.554..207M,2018ATel11245....1T}.

The late turn-on of the X-ray and radio emission from \object{GW170817} is not consistent with emission produced via an on-axis relativistic jet \citep{2017ApJ...848L..21A,2017ApJ...848L..25H,2017Sci...358.1579H,2017ApJ...848L..20M,2017Natur.551...71T}. Moreover, the gradual rise of the radio light curve rules out prompt $\gamma$-ray emission originating from a jet with a ``top-hat" azimuthal density profile observed off-axis, which would have produced a much steeper peak and decline than observed \citep{2002ApJ...570L..61G,2002ApJ...579..699N}. Instead, the light curve is consistent with mildly relativistic quasi-spherical outflow called a ``cocoon" \citep{2017Sci...358.1579H,2018MNRAS.473..576G,2018Natur.554..207M,2018arXiv180109712N} which may have some contribution from an embedded relativistic jet observed off-axis \citep[some versions of which are also referred to as a ``structured jet";][]{2017arXiv171203237L,2018arXiv180106164D,2018arXiv180103531M,2018arXiv180302768R}.

Based on the data available in the literature to date, it is not possible to establish whether or not a successful jet is present within the cocoon, as these scenarios exhibit similar behavior in the early stages of the afterglow evolution; or to determine the energy of the cocoon itself \cite[see Figure 5 of][]{2018arXiv180103531M}. The timescale of the peak flux density and the rate of decline afterwards can constrain the total energy of the outflow and the properties of a successful jet (if present). If the jet did not successfully break out of the cocoon (referred to as a ``choked'' jet) the observed emission is dominated by the quasi-spherical outflow \citep[cocoon or dynamical ejecta;][]{2018MNRAS.473..576G} and the light curve will continue to rise; if the jet is successful \citep[structured jet;][]{2018arXiv180103531M,2018arXiv180109712N} the light curve peaks sooner and declines more rapidly.  In either case identifying when and how the light curve peaks also allows calorimetry of the cocoon emission (much as was done by \citealt{2000ApJ...537..191F,2004ApJ...612..966B} for long GRBs).

To date, X-ray observations provide conflicting evidence as to whether the afterglow has peaked. \textit{XMM-Newton} observations 135 days post-merger suggest the afterglow may have flattened \citep{2018arXiv180106164D}, but \textit{Chandra} observations show a continued rise or slow turnover at about 150 days post-merger \citep{2018ATel11242....1H,2018ATel11245....1T,2018arXiv180103531M}. A decreasing X-ray brightness would imply that either the synchrotron cooling frequency has shifted below the X-ray band (expected on timescales of 100--1000 days post-merger) and the spectrum of the source has evolved, or the light curve of the source from the radio to X-rays has peaked, but current data are not definitive that \textit{any} change in the X-ray light curve has occurred.

We present further radio observations of \object{GW170817} using the the Australia Telescope Compact Array (ATCA) and the Karl G.\ Jansky Very Large Array (VLA), covering the period 125--200 days post-merger. These observations demonstrate (Figure~\ref{fig:fitting}) that the radio afterglow has peaked at $149\pm2$ days post-merger and is now declining in flux density.

\section{Observations and Data reduction} \label{sec:data}
\begin{deluxetable}{cccccccC}[b!]
\tablecaption{New radio observations of \object{GW170817} \label{tab:radiodata}}
\tablecolumns{6}
\tablenum{1}
\tablewidth{0pt}
\tablehead{
\colhead{UT date} &
\colhead{$\Delta$T} &
\colhead{Telescope} & \colhead{$\nu$} & \colhead{Bandwidth} & \colhead{Beam Size} & \colhead{$S_\nu$}\\
\colhead{} & \colhead{(d)} & \colhead{} & \colhead{(GHz)} & \colhead{(GHz)} & \colhead{(arcsec)} & \colhead{($\mu$Jy)}
}
\startdata
2017 Dec 20.83 & 125.30 & ATCA\tablenotemark{a} & 5.5 & 2.048 & 5.8$\times$1.5 & 82.0 $\pm$ 9.3\phn\\
 & &  & 9.0 & 2.048 & 3.6$\times$1.0 & 63.7 $\pm$ 8.2\phn\\
2018 Jan 13.79 & 149.26 & ATCA\tablenotemark{a} & 5.5 & 2.048 & 5.4$\times$1.5 & 98.9 $\pm$ 8.5\phn\\
& &  & 9.0 & 2.048 & 3.3$\times$1.0 & 52.7 $\pm$ 6.5\phn\\
2018 Feb 01.74 & 168.21 & ATCA\tablenotemark{b} & 5.5 & 2.048 & \nodata & \nodata\tablenotemark{c}\\
& &  & 9.0 & 2.048 & \nodata & \nodata\tablenotemark{c}\\
2018 Feb 15.17 & 181.64 & ATCA\tablenotemark{d} & 5.5 & 2.048 & 4.4$\times$1.1 & 89.6 $\pm$ 13.3\\
 & &  & 9.0 & 2.048 & 2.6$\times$0.7 & 57.0 $\pm$ 10.9\\
2018 Mar 02.32 & 196.79 & VLA\tablenotemark{e} & 2.5 & 1 & 1.3$\times$0.5 & 91.0 $\pm$ 9.1\phn\\
& &  & 3.5 & 1 & 1.3$\times$0.5 & 66.9 $\pm$ 6.1\phn\\
\enddata
\tablenotetext{a}{With the 6C configuration (maximum baselines of 6\,km) and program CX391 (PI: T.~Murphy).}
\tablenotetext{b}{With the 750A configuration (maximum baseline of 3.75\,km) and program CX394 (PI: E.~Troja).}
\tablenotetext{c}{Insufficient data quality}
\tablenotetext{d}{With the 750B configuration (maximum baseline of 4.5\,km) and program CX394 (PI: E.~Troja).}
\tablenotetext{e}{With the A configuration (maximum baseline of 27\,km) under a Director Discretionary Time program (VLA/17B-397;
PI: K. Mooley).}
\end{deluxetable}

\subsection{ATCA} \label{subsec:atca}

We observed \object{GW170817} on 2017 December 20 and 2018 January 13 UT with the ATCA (PI: Murphy). Further observations of \object{GW170817} with the ATCA were obtained on 2018 February 01 and 15 UT (PI: Troja); see Table~\ref{tab:radiodata} for details. The February 01 observation only had 4 out of 6 antennas available and after removing short baselines due to the compact configuration, the data quality was insufficient to make a meaningful measurement and the observation was discarded. We determined the flux scale and bandpass response for all epochs using the ATCA primary calibrator \object[PKS B1934-638]{PKS~B1934$-$638}. Observations of \object[PKS B1245-197]{PKS~B1245$-$197} were used to calibrate the complex gains during the December and January observing epochs, while \object[PKS B1244-255]{PKS~B1244$-$255} was used in the February observation. All observations used two bands of 2048 MHz centered at 5.5 and 9.0\,GHz. 

We reduced the visibility data using standard MIRIAD \citep{1995ASPC...77..433S} routines. The calibrated visibility data were split into the 5.5 and 9.0\,GHz bands, averaged to 32\,MHz channels, and imported into DIFMAP \citep{1997ASPC..125...77S}. Bright field sources were modeled separately for each band using the visibility data and a combination of point-source and Gaussian components with power-law spectra. After subtracting the modeled field sources from the visibility data, \object{GW170817} dominates the residual image. Restored naturally-weighted images for each band were generated by convolving the restoring beam and modeled components, adding the residual map and averaging to form a wide-band image. Image-based Gaussian fitting with unconstrained flux density and source position was performed in the region near \object{GW170817}. The resulting source position agrees with the position of \object{GW170817} observed by the \textit{Hubble Space Telescope} \citep[\textit{HST},][]{GCN21816}.

To examine the stability of the absolute flux calibration from epoch to epoch we measured the flux density of the phase calibrator (PKS~B1245$-$197) and a compact reference source in the \object{GW170817} field (RA$=13\rah09\ram53\fs91$, Dec$=-23\arcdeg21\arcmin34\farcs5$, $1.9\arcmin$ from \object{GW170817}) in each epoch and frequency band of the ATCA data. We do not use the host galaxy \object{NGC 4993} as it is extended. We find that the mean and standard deviation of the phase calibrator flux density is $2.193\pm0.013$\,Jy and $1.449\pm0.021$\,Jy at 5.5\,GHz and 9\,GHz, respectively. This compares to within $0.1\%$ with the values reported by the ATNF Calibrator Database\footnote{\url{http://www.narrabri.atnf.csiro.au/calibrators/}}.
The reference source is three orders of magnitude fainter than the phase calibrator but is a factor of at least three brighter than \object{GW170817} and is within the same field, so it should provide an accurate indication of the flux density scale within the target field itself. The source is also visible regardless of which phase calibrator is used and so provides an independent test of flux scale stability. Across all epochs, we find that the mean flux density and standard deviation of the reference source flux density is $452\pm16\,\mu$Jy and $301\pm18\,\mu$Jy at 5.5\,GHz, and 9.0\,GHz, respectively. This suggests that our field flux density measurements are stable to within $2.9\%$ and $5.4\%$ at 5.5\,GHz and 9.0\,GHz, respectively, where those additional uncertainties when added in quadrature to the measurement uncertainties give reduced $\chi^2 = 1$ for the reference source. For \object{GW170817} itself we measured the noise in the vicinity of the source to account for additional contributions from unmodeled sidelobes from the host galaxy \object[NGC 4993]{NGC~4993} and included the additional uncertainties discussed above.

\subsection{VLA} \label{subsec:vla}
VLA observations of the \object{GW170817} field were carried out on 2018 March 02 (Table~\ref{tab:radiodata}). The Wideband Interferometric Digital Architecture (WIDAR) correlator was used at S band (2--4\,GHz) to maximize sensitivity. We used \object{J1248-1959} as the phase calibrator and \object{3C286} as the flux density and bandpass calibrator. The data were calibrated and flagged for RFI using the NRAO {CASA} \citep{2007ASPC..376..127M} pipeline. We then split and imaged the target data using the CASA tasks {\tt split} and {\tt clean}. We made final images by splitting the bandpass into 2 subbands of 1\,GHz each.

\begin{figure*}
\plotone{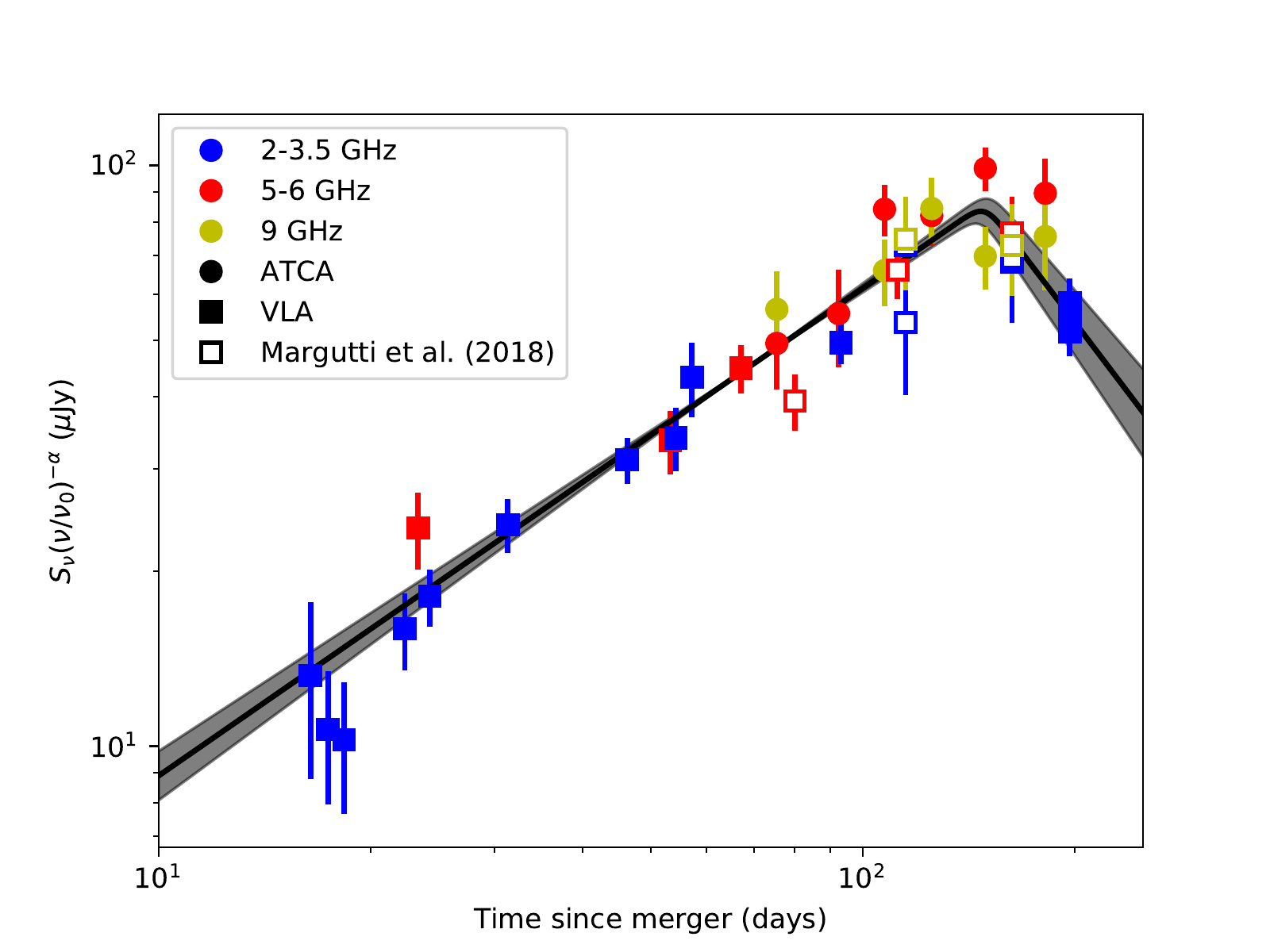}
\caption{Light curve of \object{GW170817} from ATCA (circles) and VLA (squares) observations grouped by frequency band, with 2--3.5\,GHz (blue), 5--6\,GHz (red), and 9\,GHz (yellow). The flux densities have been adjusted to 5.5\,GHz assuming a spectral index of $\alpha=-0.57\pm 0.04$ (\S~\ref{sec:spectral}). Open squares denote observations from \citet{2018arXiv180103531M}, while filled symbols denote observations from this paper or other observations by our group \citep{2017Sci...358.1579H,2018Natur.554..207M}. Our best-fit smoothed broken power-law with temporal index on the rise $\delta_1=0.84\pm 0.05$, temporal index on the decay $\delta_2=-1.6\pm0.2$ and peak time $t_{\rm peak}=149\pm 2$\,days is shown in black, with uncertainties shaded.\label{fig:fitting}}
\end{figure*}

\section{Results and Discussion} \label{sec:results}

\begin{figure}
\plotone{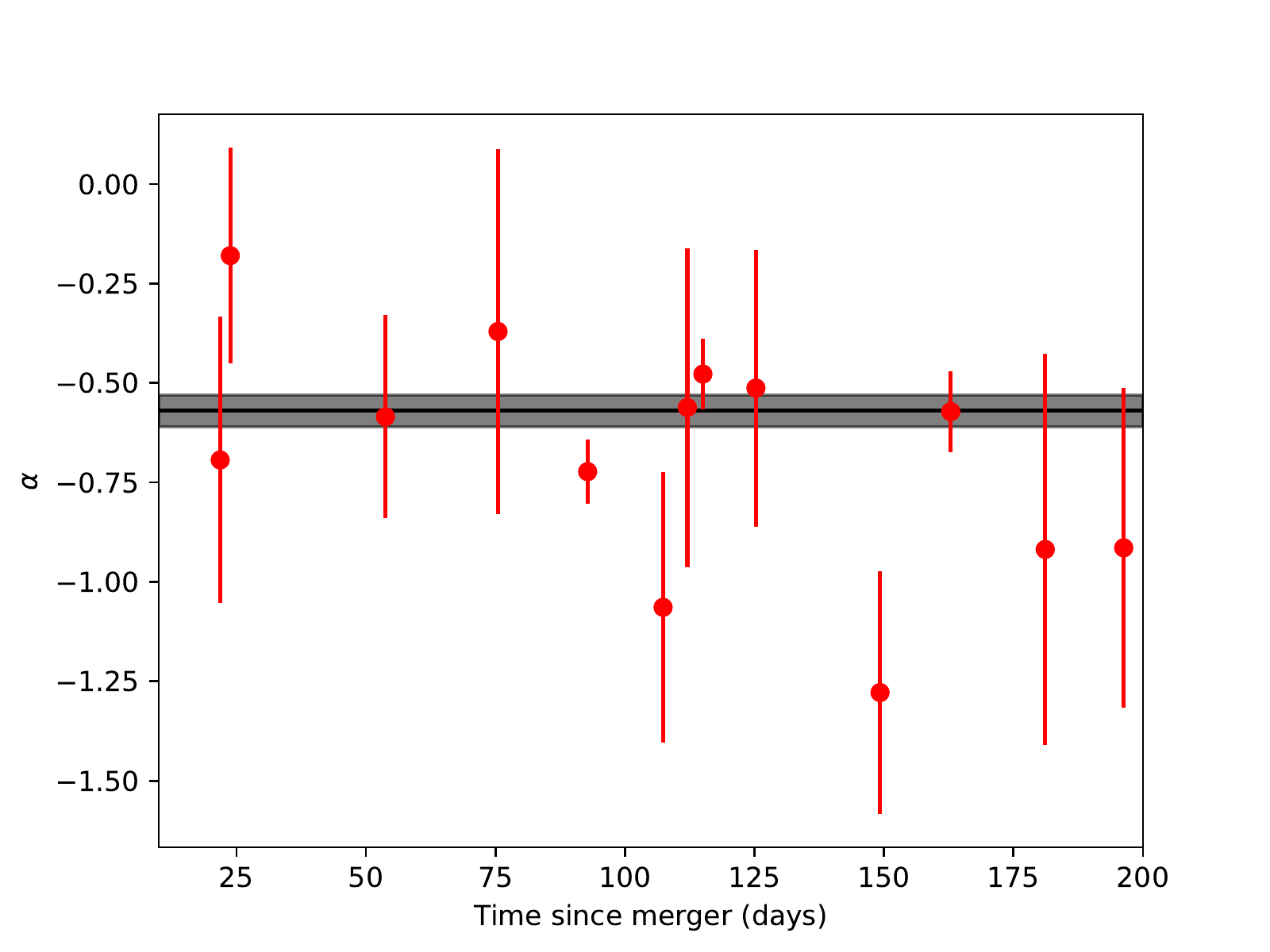}
\caption{Spectral index ($\alpha$) of contemporaneous radio observations from \citet{2017Sci...358.1579H}, \citet{2018Natur.554..207M}, \citet{2018arXiv180103531M} and this work. The best-fit spectral index from the first 120 days of the radio light curve, $\alpha=-0.57\pm 0.04$, is shown in black, with uncertainties shaded.\label{fig:alpha_timeseries}}
\end{figure}

\subsection{Spectral analysis}
\label{sec:spectral}
We first re-visit the spectral behavior of the radio emission.
As in \citet{2018Natur.554..207M} we fit a power-law of the form $S_\nu \propto \nu^{\alpha}t^{\delta}$ to the first 120 days of the radio light curve (before any sign of a turnover) and find a spectral index $\alpha=-0.57\pm 0.04$ and temporal index $\delta=0.84\pm0.05$. This is consistent with \citet{2018Natur.554..207M} and with \citet{2018arXiv180103531M}, who find a joint radio-to-X-ray spectral index $\alpha=-0.585\pm0.005$ at 110 days and $\alpha=-0.584\pm0.006$ at 160 days post-merger.

We examined the variability of the spectral behavior using all quasi-simultaneous radio observations.  We identified data-sets with more than one observation within $\pm1\,$day and fit for a spectral index. These values are shown in Figure~\ref{fig:alpha_timeseries}. We find the data largely consistent with a constant spectral index, with $\chi^2=15.9$ for 12 degrees-of-freedom.  There appears to be no evidence for significant change in the spectrum of the source, consistent with previous radio, X-ray and \textit{HST} observations \citep{2018arXiv180106164D,2018arXiv180102669L,2018Natur.554..207M,2018arXiv180103531M,2018arXiv180302768R}.

\subsection{Light curve analysis} \label{subsec:light curve}
Figure \ref{fig:fitting} shows the light curve of \object{GW170817} over the 2--9\,GHz frequency range from the observations in Table~\ref{tab:radiodata} and the literature \citep{2017Sci...358.1579H,2018Natur.554..207M,2018arXiv180103531M}, scaling the flux density for each observation to 5.5\,GHz based on the spectral index of $\alpha=-0.57\pm 0.04$ calculated above. Assuming the light curve initially rises with a temporal index of $\delta_1=0.84$, peaks $t_{\rm peak}$\,days post-merger, and fades with a temporal index of $\delta_2$, we fit a smoothed broken power law\footnote{\url{http://docs.astropy.org/en/stable/api/astropy.modeling.powerlaws.SmoothlyBrokenPowerLaw1D.html}} using the \texttt{Astropy} modeling package \citep{2018arXiv180102634T} that behaves as $S_\nu\propto t^{\delta_1}$ for $t\lesssim t_{\rm peak}$ and $S_\nu\propto t^{\delta_2}$ for $t\gtrsim t_{\rm peak}$ with a smooth transition around $t\approx t_{\rm peak}$. We do not expect to see any variability due to interstellar scintillation, due to the source size \citep{2017Sci...358.1579H}.

We have fit the lightcurve allowing the smoothing factor to freely vary and find a minor preference for small smoothing factors down to 0.001, corresponding to a transition of 0.3 days either side of the break. To approximate our observing cadence near the peak of the lightcurve we use a smoothing factor of 0.02 (corresponding to a $<20$ day transition) which produces no significant changes in fit parameters.

\begin{figure}
\plotone{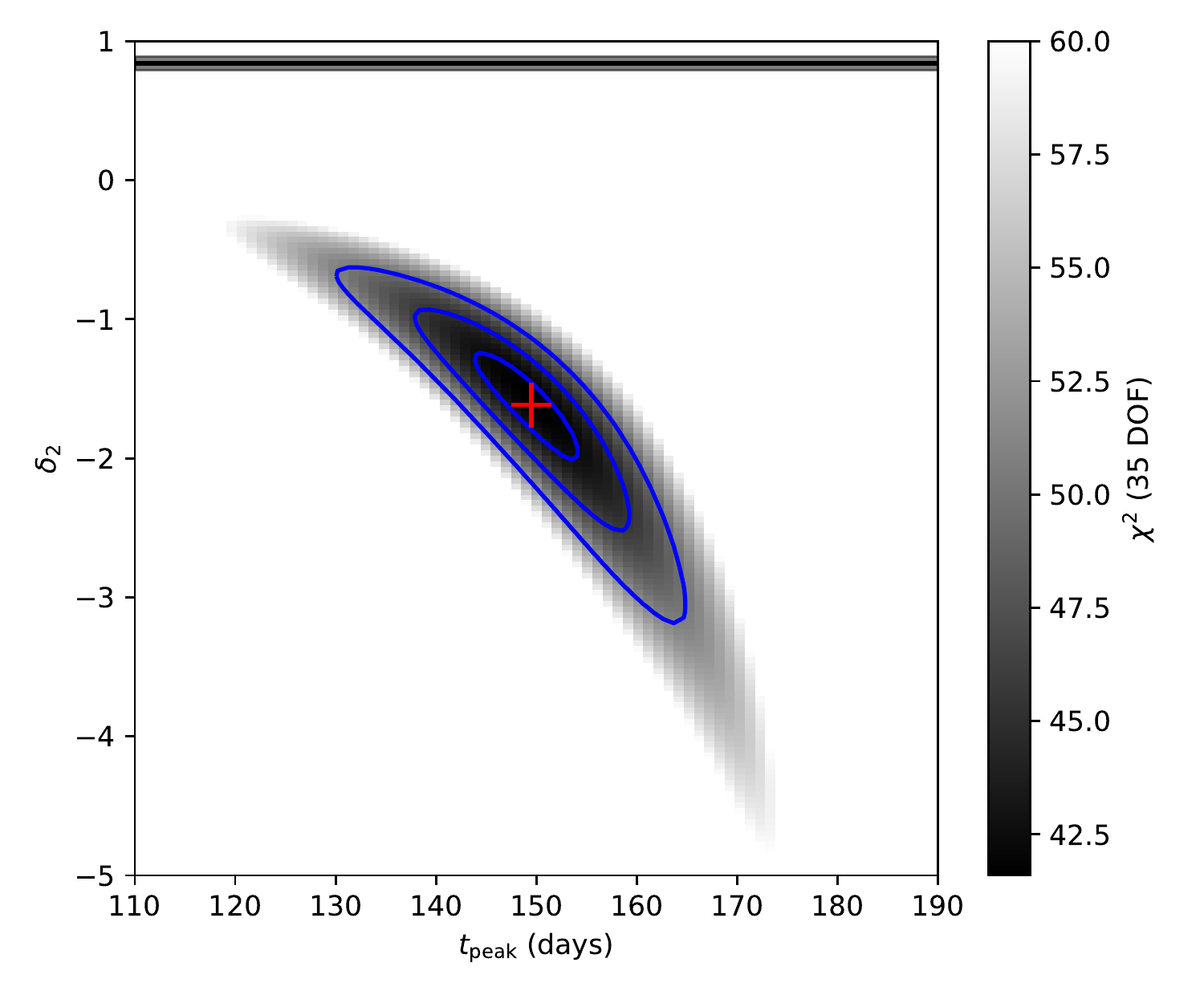}
\caption{Two-dimensional joint probability distribution of $\delta_2$ and $t_{\rm peak}$, assuming $\alpha=-0.57\pm 0.04$ and $\delta_1=0.84$. The background gray-scale is the $\chi^2$ for 35 degrees-of-freedom, with  1-, 2-, and 3-$\sigma$ joint confidence contours are shown in blue. The best-fit value of $\delta_2=-1.6\pm0.2$ and $t_{\rm peak}=149\pm2\,$days is shown in red.  The temporal index of the light curve as it rises, $\delta_1=0.84\pm0.05$, is indicated by the black line with uncertainties shaded. \label{fig:joint_distribution}}
\end{figure}

Figure \ref{fig:joint_distribution} shows the two dimensional joint confidence region as a function of $t_{\rm peak}$ and $\delta_2$, where we indicate the best-fit values, $\delta_2=-1.6 \pm 0.2$ and $t_{\rm peak}=149 \pm 2$\,days, and the 90\% confidence region. The best fit has $\chi^2=41.6$ for 35 degrees-of-freedom. For a radio light curve that is continuing to rise, the temporal index would remain the same, $\delta_2=\delta_1$, which we indicate with the dashed line in Figure~\ref{fig:joint_distribution}. Comparing the $\chi^2(\delta_2=\delta_1)$ to the minimum $\chi^2$ for $\delta_2=-1.6$, we find a change of 380 for one additional parameter and can exclude a light curve that continues to rise at greater than 5$\sigma$ significance using an F-test. We further find a change of $\chi^2$ of 35 from $\delta_2=0$ to the best-fit value $\delta_2=-1.6\pm0.2$, leading to a declining light curve. Preliminary reduction of further observations confirms the observed trend.

\subsection{Interpreting the radio light curve} \label{subsec:models}
The observed light curve turns over and declines with no evidence for a steep rise coming with an energetically-dominant off-axis jet \citep{2018arXiv180109712N}, but a weaker jet may still be present. The relatively sharp peak in the radio light curve implies that the energy injection has reduced substantially (or stopped), or that the ejecta has collected mass comparable to its own. The former scenario would be relevant for a successful jet \citep[e.g.,][]{2017Sci...358.1559K,2017arXiv171203237L,2018Natur.554..207M,2018arXiv180103531M,2018arXiv180106516T,2018arXiv180106164D} or a low energy choked-jet cocoon \citep[e.g.,][]{2017Sci...358.1559K,2017arXiv171005896G,2018ApJ...855..103P,2018Natur.554..207M}, while the latter would be relevant in the case of an isotropic fireball \citep[i.e., dynamical ejecta;][]{2011Natur.478...82N,2018Natur.554..207M,2018arXiv180106164D,2018arXiv180300599H}.

While no substantial degree of linear polarization would be expected from isotropic dynamical ejecta, in the successful jet model the required asymmetry is built into the jet structure \citep[the energy and speed of the various ejecta components are both functions of the angle from the jet axis; see e.g.][]{2017arXiv171203237L}. Thus, the relevant emitting surface is never completely symmetric for misaligned observers, resulting in an appreciable degree of linear polarization \citep[$\sim 20\%$;][]{2004MNRAS.354...86R}. A detection of significant linear polarization would thus point to a successful jet rather than  isotropic dynamical ejecta \citep[also see][]{2018arXiv180305892G}.

The radio light curve can give the energy profile of the ejecta, but it is not sufficient for distinguishing between the contributions from radial and angular structures within the ejecta. Very Long Baseline Interferometry (VLBI) can, however, provide images at sub-milliarcsecond angular resolution, and thus constrain the geometry of the outflow. Distinguishing between the successful-jet, choked-jet cocoon and dynamical ejecta models is thus possible using VLBI observations.

The time of the radio peak is near the observed plateau on the X-ray light curve \citep{2018arXiv180103531M,2018arXiv180106516T,2018ApJ...853L...4R,2018arXiv180106164D}, and suggests that the X-rays peaked at the same time as the radio light curve.
The turnover in the X-ray (and radio) light curve is therefore dynamical or geometric in origin, and the cooling break has (likely) not entered the X-ray band yet.
This is consistent with the interpretation of \cite{2018arXiv180106164D} and \cite{2018arXiv180103531M} who find that the radio, optical and X-rays lie on the same power-law  until day 150 post-merger.

The light curve of a relativistic jet afterglow will decay as $t^{-p}$, while in the non-relativistic regime the decline will be proportional to $t^{(15p-21)/10}$, with $p$ the exponent on the distribution of electron energies, $N(E) \propto E^{-p}$ \citep{2002ApJ...570L..61G,2011Natur.478...82N}. In the case of \object{GW170817}, $p=2.17$ \citep[e.g.][]{2018arXiv180103531M,2018Natur.554..207M}, so the expected decay slopes are $t^{-2.2}$ and $t^{-1.2}$. Our radio data are consistent with expectations for the mildly- or non-relativistic regimes. Based on the time and the flux density at the peak of the radio light curve, we can further calculate the isotropic equivalent energy \citep{2018arXiv180109712N} as a few $\times$ $10^{50}$\,erg for the cocoon scenario (also see \citealt{2018arXiv180302768R})
and a few $\times$ $10^{49}$\,erg for the dynamical ejecta scenario.  Both of those are  lower than the isotropic-equivalent kinetic energies found for short GRBs \citep{2015ApJ...815..102F}.

If the peak of the light curve was dominated by an off-axis jet, then $\Gamma(\theta_{\rm obs}-\theta_{\rm jet})\simeq1$ \citep[where the bulk Lorentz factor of the jet is $\Gamma$, the off-axis angle of the observer is $\theta_{\rm obs}$, and the opening angle of the jet is $\theta_{\rm jet}$]{2018arXiv180109712N} implies that $(\theta_{\rm obs}-\theta_{\rm jet})\simeq20\degr$, assuming that material with $\Gamma\simeq3$ dominated the on-axis emission at peak. Therefore we can constrain $\theta_{\rm jet}\lesssim 8\arcdeg$ using the viewing angle constraint from LIGO/Virgo \citep[$\theta_{\rm obs}<28\arcdeg$;][]{2017PhRvL.119p1101A}. 

Continued radio monitoring will be essential for constraining the decay index. A steep decline in the radio light curve would favor the scenario in which a successful jet broke out of the dynamical ejecta. Transition of the ejecta from the mildly-relativistic to the Newtonian regime would be characterized by deviation from a power-law decay and a change in spectral index, which could be detected with sensitive follow up observations. It is even possible for the ejecta to have angular structures that could cause the light curve to rise again: the early-time kilonova signal in the optical suggested the presence of $\sim$ 0.05 M${_\odot}$ material traveling at speeds of 0.1c to 0.3c which should give rise to a radio peak on timescales of a few years \citep{2017ApJ...848L..21A,2011Natur.478...82N,2018arXiv180109712N}. Finally, the full radio light curve of \object{GW170817} will be crucial for calorimetry, since it will capture all of the energy in the ejecta. The total energy will further shed light into whether \object{GW170817} is a standard short GRB viewed off-axis or it represents a distinct phenomenon.

\section{Conclusion}
We have presented new ATCA and VLA observations of \object{GW170817} covering the period 125--200\,days post-merger. Combined with previous radio observations these data show no evidence for spectral evolution, but they conclusively show that the radio counterpart has peaked in brightness at $149\pm2$\,days post-merger and is currently declining.  We use this to rule out emission being caused by highly energetic, quasi-isotropic outflow or highly energetic, highly-relativistic outflow  but are not able to uniquely determine the geometry and structure of the actual outflow material. Continued radio monitoring will allow the temporal decay index to be accurately determined, although this may not be sufficient to establish the presence of a successful jet \citep{2018arXiv180109712N} and degeneracies in the ejecta total energy and the density of the circum-merger environment may preclude confirmation of any particular model. Polarization measurements and VLBI observations should be able to break this degeneracy and thus distinguish between the models \citep[also see][]{2018arXiv180305892G}.

\acknowledgements
We thank P.~Chang for helpful discussions.  TM acknowledges the support of the Australian Research Council through grant FT150100099. AC acknowledges support from the NSF CAREER award \#1455090. 
Part of this research was conducted by the Australian Research Council Centre of Excellence for All-sky Astrophysics (CAASTRO), through project number CE110001020. 
Part of this work was supported by the GROWTH (Global Relay of Observatories Watching Transients Happen) project funded by the National Science Foundation under PIRE Grant No 1545949. DK was additionally supported by by NSF grant AST-1412421. The National Radio Astronomy Observatory is a facility of the National Science Foundation operated under cooperative agreement by Associated Universities, Inc. K.P.M. is currently a Jansky Fellow of the National Radio Astronomy Observatory. The Australia Telescope is funded by the Commonwealth of Australia for operation as a National Facility managed by CSIRO.

\facility{ATCA, VLA}
\software{Astropy \citep{2018arXiv180102634T}, MIRIAD \citep{1995ASPC...77..433S}, DIFMAP \citep{1997ASPC..125...77S}, CASA \citep{2007ASPC..376..127M}}.


\end{document}